\documentclass[aps,prb,twocolumn,showpacs,superscriptaddress]{revtex4}
\usepackage{amsfonts}
\usepackage{amsmath}
\usepackage{amssymb}
\usepackage{graphicx}

\begin{document}

\preprint{}
\title{Manipulation of Majorana states in X-junction geometries}
\author{D. N. Aristov}
\affiliation{Petersburg Nuclear Physics Institute, NRC ``Kurchatov Institute'', Gatchina 188300, Russia}
\affiliation{Department of Physics, St.Petersburg State University, Ulianovskaya 1,
St.Petersburg 198504, Russia}
\affiliation{Institute for Nanotechnology, Karlsruhe Institute of Technology, 76021
Karlsruhe, Germany }
\author{D. B. Gutman}
\affiliation{ Department of Physics, Bar Ilan University, Ramat Gan 52900, Israel}
\keywords{one two three}
\pacs{03.67.Lx,71.10.Pm,03.65.Vf}

\begin{abstract}
We study quantum manipulation based on four Majorana bound states  in  X-junction geometry. 
The  parameter space of this setup  is bigger than of  the previously studied Y-junction  and is 
described by SO(4) symmetry group. In order for quantum computation to be dephasing free,   
two Majorana states have to stay degenerate at all times.
We find a  condition necessary for  that and compute  the Berry's phase, $2\alpha$, accumulated during the manipulation.  We construct  simple protocols for  the variety of values of $\alpha$, 
including $\pi/8$ needed for the purposes of quantum computation.
Although the manipulations in general X-junction geometry  are not topologically protected, 
they may prove to be a feasible compromise for aims of quantum computation.
\end{abstract}

\volumeyear{year}
\volumenumber{number}
\issuenumber{number}
\eid{identifier}
\date{\today}
\maketitle

 
 
\section{Introduction}
Quantum computing has a huge advantage over a classical one for a simulation of   physical experiments, 
as well  as for implementation of certain algorithms \cite{Nielsen}.  Significant efforts were invested in  building the quantum computer.  Its  solid state realizations   are   limited by  dephasing processes that destroy the coherence,  and ruin the computation. Topologicaly protected quantum computations is a promising way to  overcome this problem.\cite{Kitaev,Ivanov}  
The latter  employ  non-Abelian states,  that are  not local and  are not affected by an environment. \cite{Beenakker2013,Stern2013}
The simplest realization  of these states are Majorana fermions.
Since the early proposals a large number of solid state realizations  were  envisioned, 
based on topological insulators \cite{Fu2008,Linder2010}, semiconductor heterostructures
\cite{Sau2010,Alicea2010}, noncentrosymmetric
superconductors\cite{Sato2009}, and quantum Hall systems at integer plateau transitions\cite{Lee2009},
as well as  one-dimensional  semiconducting wires deposited on an s-wave superconductor \cite{Lutchin2010,Oreg2010}.  The signatures of Majorana states were detected  in the recent experiments\cite{Kouwenhoven2012,Heiblum2012,Xu2012,Marcus2013,Xu2014}.

Besides having the protected memory units (q-bits), the  implementation of quantum computers  requires 
an initialization,  manipulation and read out. These processes involve 
unitary transformations of the degenerate ground states achieved by braiding of Majorana fermions.
For one dimensional   realization of Majorana modes  the braiding  is achieved   by building a network of wires\cite{Alicea2011,Sau2011}. These wires are  connected by Y-junctions (see Fig. \ref{fig:Yjunction}), that   are  controlled  by the  external gates \cite{Halperin2012},
supercurrents \cite{Romito2012},  or magnetic fluxes\cite{Hyart2013}.

Unfortunately,   the set of existing (albeit in theory) gates is  incomplete, 
and can not be completed using  Majorana fermions alone.
The conventional proposals lack the so-called  $\pi/8$ gate, 
so that  the  desired topological computation remains elusive.
The existing constructions for the latter gate rely on the non-protected manipulations,   
with the error correction protocol\cite{Karzig2015}.
Since the complete dephasing-proof realization  is  not yet found,  it makes sense to consider 
a different compromise. In the present  work we discuss the manipulation  via
X-junctions, schematically shown in Fig.\ref{fig:Xjunction}. Such junctions naturaly arise in 
the circuit theory of quantum computing and are realized experimentally \cite{Plissard2013}.  

The parameter space of  the underlying Hamiltonian with four localized Majorana states, is considerably larger than the  previously studied  case of Y-junction. This allows us to implement additional gates at the price of  dealing with the general form of interaction between four  Majorana fermions.  
Generally, any  state with an even number of Majorana fermions has no topological protection. 
Nevertheless,  we show below that the parameters of X-junction can  be specially tuned in 
such a way that the ground state remains degenerate   and the states  are not affected by dephasing.
It means that mastering a good control of  the junction amounts  to reducing 
the dephasing level to an arbitrary low level.  We hope, that this may be a  practical way to implement dephasing-proof quantum computation.
 
\section{Y-junction}

\begin{figure}[t]
\includegraphics[width=.8\columnwidth]{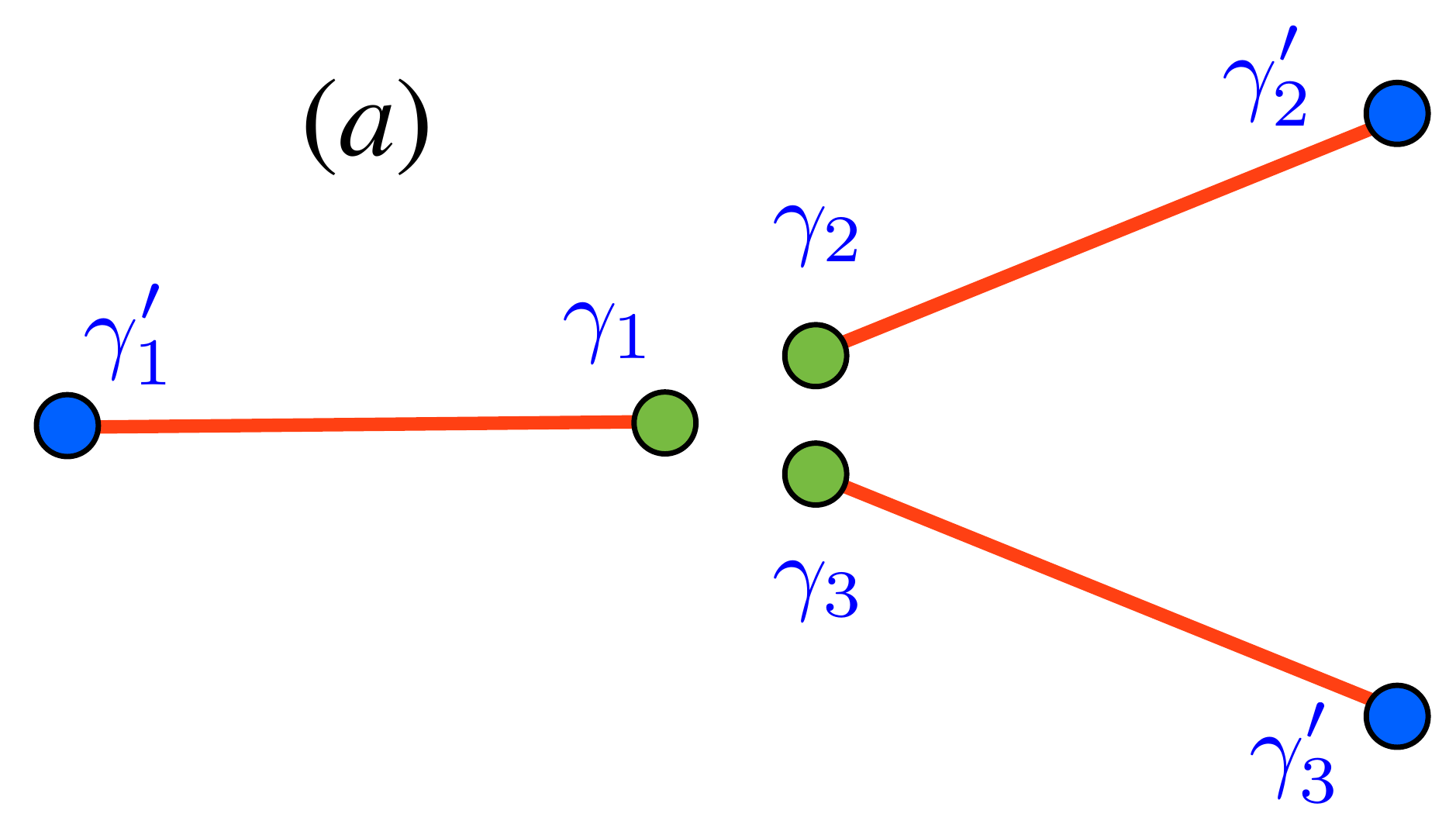} 
\includegraphics[width=.8\columnwidth]{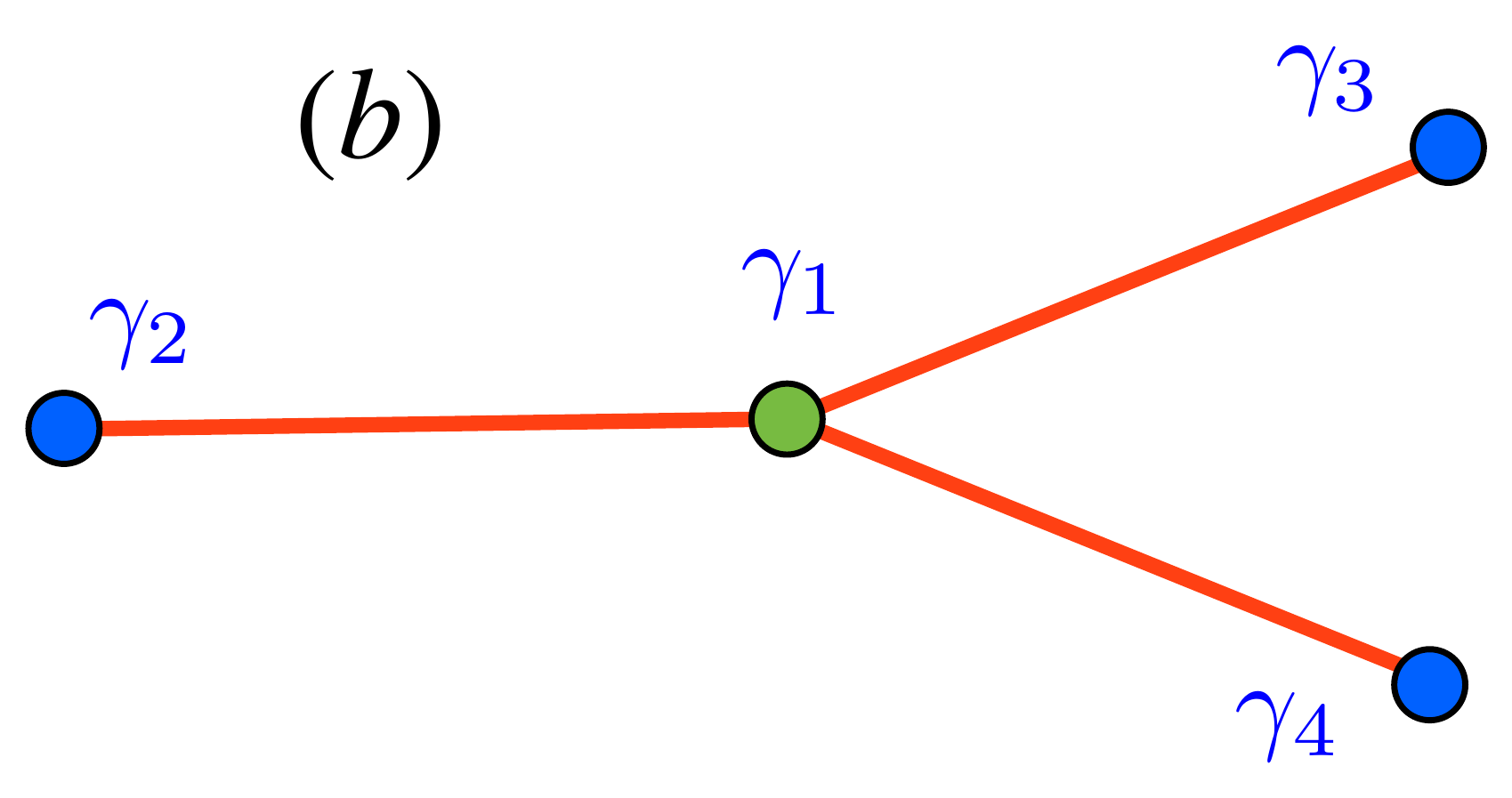} 
\caption{\label{fig:Yjunction}  The usually discussed  Y-junction setup. The initial configuration is shown in   panel (a) and consists of  three three Majorana states, $\gamma_{i}$, in the center and other three states, $\gamma_{i}'$, at some distance. After hybridization, two out of three central Majorana states are  split and the third one remains at zero energy. The resulting configuration is shown after renumbering in panel (b) and is described by the effective Hamiltonian  \eqref{eq:Yjunction}.  
  }
\end{figure}

We start our discussion with  a Y-junction geometry, shown in Fig. \ref{fig:Yjunction}a, 
that connects  three wires with the endpoint Majorana states. In this case the strongest interaction is for the central three  Majoranas $\gamma_{k}$ with $k=1,2,3$ and is given by the Hamiltonian ${\cal H}_{0} = \frac i2 \sum_{jkl}\epsilon_{jkl} h_{j} \gamma_{k}\gamma_{l} $, with  $\epsilon_{ijk}$ totally antisymmetric tensor. This Hamiltonian leads to a splitting of two Majoranas to finite energies, $E=\pm \sqrt{h_{1}^{2}+h_{2}^{2}+h_{3}^{2}}$.  One linear combination of $\gamma_{1,2,3}$ remains at zero energy, i.e. is true Majorana state. After the exclusion of irrelevant states with finite (high) energy and due renumbering, this setup is reduced to Y-junction shown in Fig. \ref{fig:Yjunction}b,  and is described by the low-energy effective Hamiltonian 
 where only one central Majorana fermion is coupled to three others.
The  Hamiltonian describing the hopping between four Majoranas $\gamma_{k}$, ($k=1,\ldots 4$) 
  \begin{equation}
 {\cal H} = 2 i \gamma_{1}\sum_{k=2}^{4} u_{k-1}  \gamma_{k}
 \label{eq:Yjunction}
 \end{equation} 
extensively studied previously, see e.g.  Refs. \cite{Beenakker2013,Stern2013} and  
the references therein.

We now parametrize this Hamiltonian by
\begin{equation}
u_{1}  = u \cos \theta , \quad   
u_{2}   = u \sin \theta \cos \phi , \quad  
u_{3}   = u \sin \theta \sin \phi .
\label{eq:param}
\end{equation}
The parameter $u$  sets an overall scale, that  is  not important for our discussion,
and  a pair of angles $(\theta,\phi)$ represents a point on a unit sphere. 
The adiabatic evolution of $ {\cal H} $ is viewed as a route, passed by this point on the sphere during the manipulation protocol. 

The above Hamiltonian has two true (non-split) Majorana eigenstates,  $\gamma_{\theta}$ and $\gamma_{\phi}$ (see below); we are following the notations of Ref.  \cite{Karzig2015}   It is convenient to combine these two states into one complex fermion $c=\gamma_{\theta} + i \gamma_{\phi}$.
During an adiabatic evolution it acquires a Berry's phase $c \to e^{i\alpha} c$,
given by 
\begin{equation}
 2\alpha  =   
 -  \oint \cos \theta\, {d\phi}  = \int \sin \theta \,d\theta d\phi  \,.
 \label{eq:Berry1}  
\end{equation}

\begin{figure}[t]
\includegraphics[width=.8\linewidth]{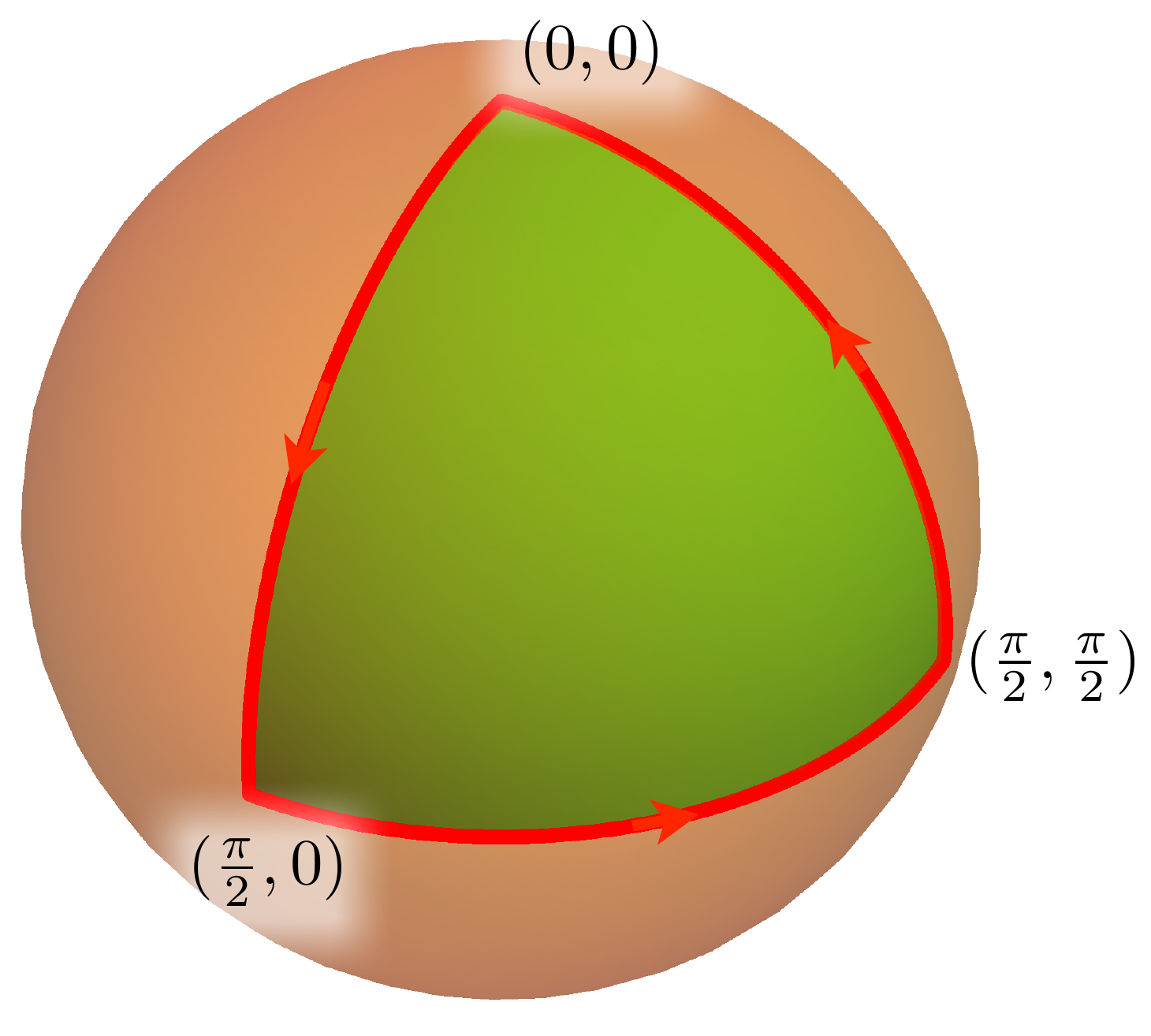} 
\caption{\label{fig:sphere}  
Unit sphere, representing the arbitrary Hamiltonian with four Majoranas. In simpler case of Y-junction the Hamiltonian \eqref{eq:Yjunction} is parametrized by the angle on the sphere $(\theta,\phi)$, and the evolution along the red line leads to $\pi/4$ Berry's phase. The case of X-junction, Eq. \eqref{eq:ham} is characterized by two points on the sphere,  $(\theta,\phi)$,  $(\bar \theta, \bar \phi)$, which parametrize the Hamiltonian with two non-split Majorana states. Possible trajectories leading to $\pi/8$ Berry's phase are discussed in the main text, one of them given by the same red line now describing the variation in $\theta$, $\phi$ while $\bar\phi$ is fixed.  
 }
\end{figure}

The simplest trajectory of adiabatic evolution corresponds to the case where one of the three parameters $u_{j}$ in Eq. \eqref{eq:Yjunction}  is zero.  
This  relies on the exponentially small overlap between wave functions of Majorana fermions  situated 
at  spatially separated  end points  of the wires.
Geometrically,  it is depicted  by three lines on the sphere, $\theta=\pi/2$, $\phi=\pi/2$ and $\phi=0$,  shown in Fig. \ref{fig:sphere}.  To simplify the presentation, let us choose the initial conditions 
such that  two computational Majoranas, $\gamma_{3,4}$, are decoupled from the ancilla Majoranas, $\gamma_{1,2}$, and from each other  ($u_{1} \neq 0$ and  $u_{2,3}  = 0$).
This corresponds to the  north pole in terms of the angular coordinates 
($\theta   = 0$, for any  value of  $\phi$). 

The evolution, that follows the 
trajectory  $(0,0) \rightarrow (\pi/2,0) \rightarrow (\pi/2,\pi/2)\rightarrow(0,0)$ 
(represented  in  Fig. \ref{fig:sphere} by a red line )
encircles the solid angle $\pi/2$. As  can be easily seen from the figure   it encompasses    one eighth of an entire sphere and is thus one eighth of its solid angle  $4\pi$.  After following this trajectory  the  wave function acquires  the  Berry's phase  $\pi/4$ that  corresponds to the interchange of two ``computational'' Majoranas, $\gamma_{4} \to \pm\gamma_{3}$, $\gamma_{3} \to \mp \gamma_{4}$.    
 Note, that this trajectory is special, because  it satisfies the condition on the angles: either  $\phi=0$ or $\theta=\pi/2$. This guarantees that Majorana states remain degenerate and the ground state  is not affected by dephasing. 
If the trajectory deviates from this line and enters the area shown in green,  the topological protection 
is lost, see below.

\begin{figure}[t]
\includegraphics[width=.8\linewidth]{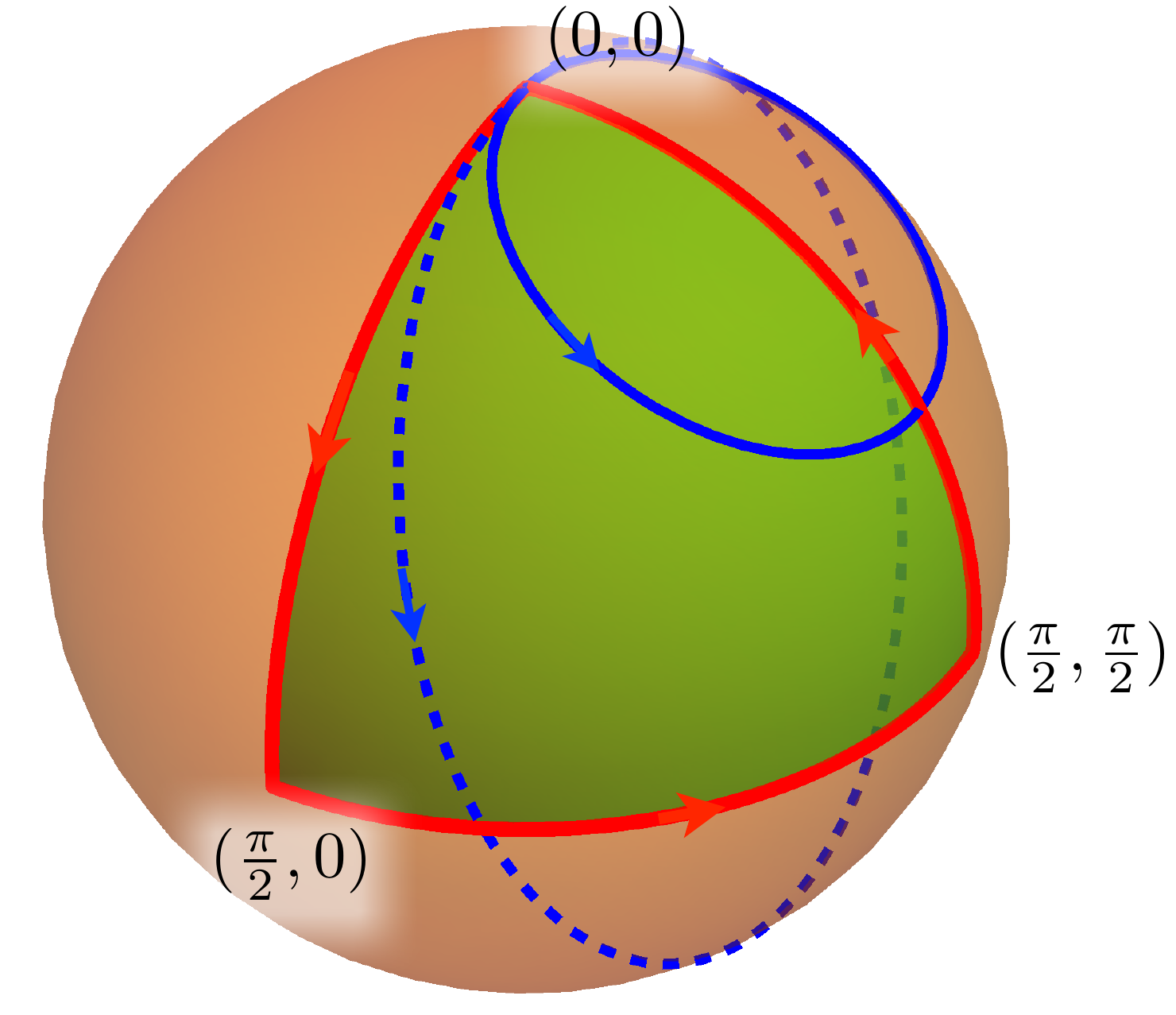} 
\caption{\label{fig:sphere2}  
Two possible trajectories in case of Y-junction, leading to $\pi/8$ Berry's phase, are  discussed in text. The first one (solid blue line) is a circle centered at  $\cos\theta_{0} =  7/8$, leading to $\alpha = \pi/8$.  Another circle is shown by dashed blue line, also passes via the north pole and is centered at $\cos\theta_{0} =  3/8$, it produces  $\alpha = 5\pi/8$ .  
 }
\end{figure}

We note here in passing, that if no concern for the topological protection were involved, the whole family of simple trajectories would provide the desired value of he Berry's phase. 
Explicitly we write in terms of  \eqref{eq:Yjunction}, 
\begin{equation}
\begin{aligned}
u_{1} & =  1-\sin^{2}\phi \tan^{2} \theta_{0} , \quad   
u_{2}   = 2\sin^{2}\phi \tan \theta_{0} , \\  
u_{3}   &= \sin2\phi \tan \theta_{0} .
\end{aligned}
\label{eq:circles}
\end{equation}
The value  $u = \sqrt{u_{1}^{2}+u_{2}^{2}+u_{3}^{2}}= 1+\sin^{2}\phi \tan^{2} \theta_{0}$ depends on $\phi$, but it is irrelevant here.  It can be easily verified that upon the variation $\phi\in (0,\pi)$ the phase $ 2\alpha = 2\pi(1-\cos\theta_{0})$, which shows that for $\cos \theta_{0} = 7/8,5/8,3/8,1/8$ we should have $ \alpha = \pi/8, 3\pi/8 ,  5 \pi/8, 7 \pi/8 $, respectively. Two examples of such circular trajectories are shown in Fig. \ref{fig:sphere2}. The apparent advantage of the parametrization \eqref{eq:circles} is the absence of sharp turning points during the adiabatic evolution (cf.\ \cite{Karzig2015}) which is characterized only by the harmonics $e^{\pm 2i\phi}$. 

It should be stressed  that  there is no way to realize $\pi/8$ state within this setup  in a topologically protected manner, and one inevitably  ends up  in   a  situation in which  all three  tunneling amplitudes are finite
($u_{j}\neq 0$) \cite{Karzig2015}.  
The sizable hopping $u_{j}$ between $\gamma$'s appears if  the distance between them is shorter that the characteristic correlation length.  
If Majorana fermions $\gamma_{2,3,4}$ are all close to $\gamma_{1}$ it implies that the distance between $\gamma_{2}$ and  $\gamma_{3}$ is of the same order.  Therefore one cannot neglect  the coupling between $\gamma_{2}$ and $\gamma_{3}$,  which is absent in Eq. \eqref{eq:Yjunction}. 
In this case we generally expect that the hopping amplitudes between all $\gamma$'s  
are non-zero.  This impedes  the quantum computation,  because the degeneracy of  the ground state has been  removed, unless the coupling constants are tuned in a special way.
Now on we focus on this case.


\section{X-junction}
\begin{figure}[t]
\includegraphics[width=.8\columnwidth]{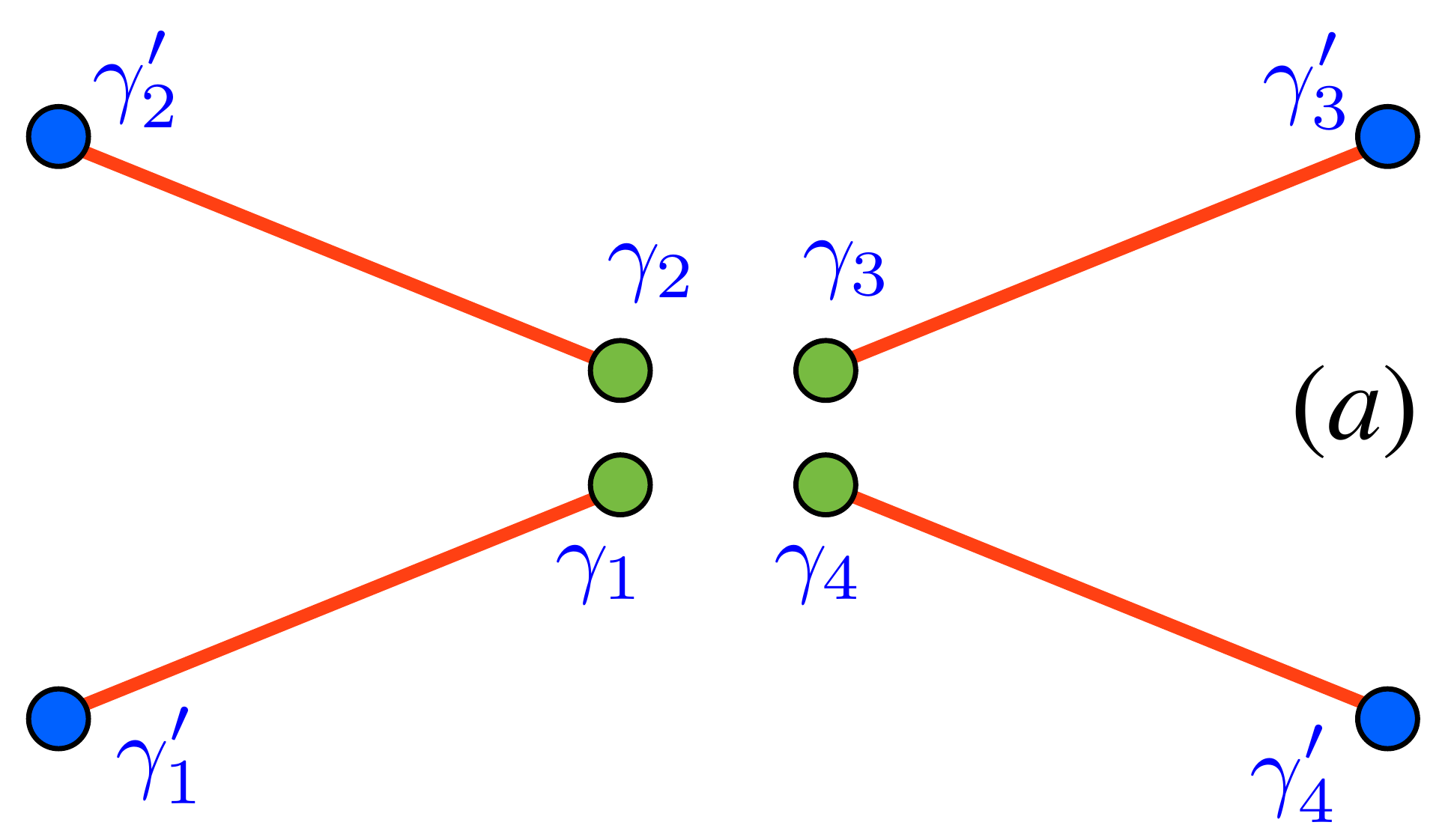} 
\includegraphics[width=.8\columnwidth]{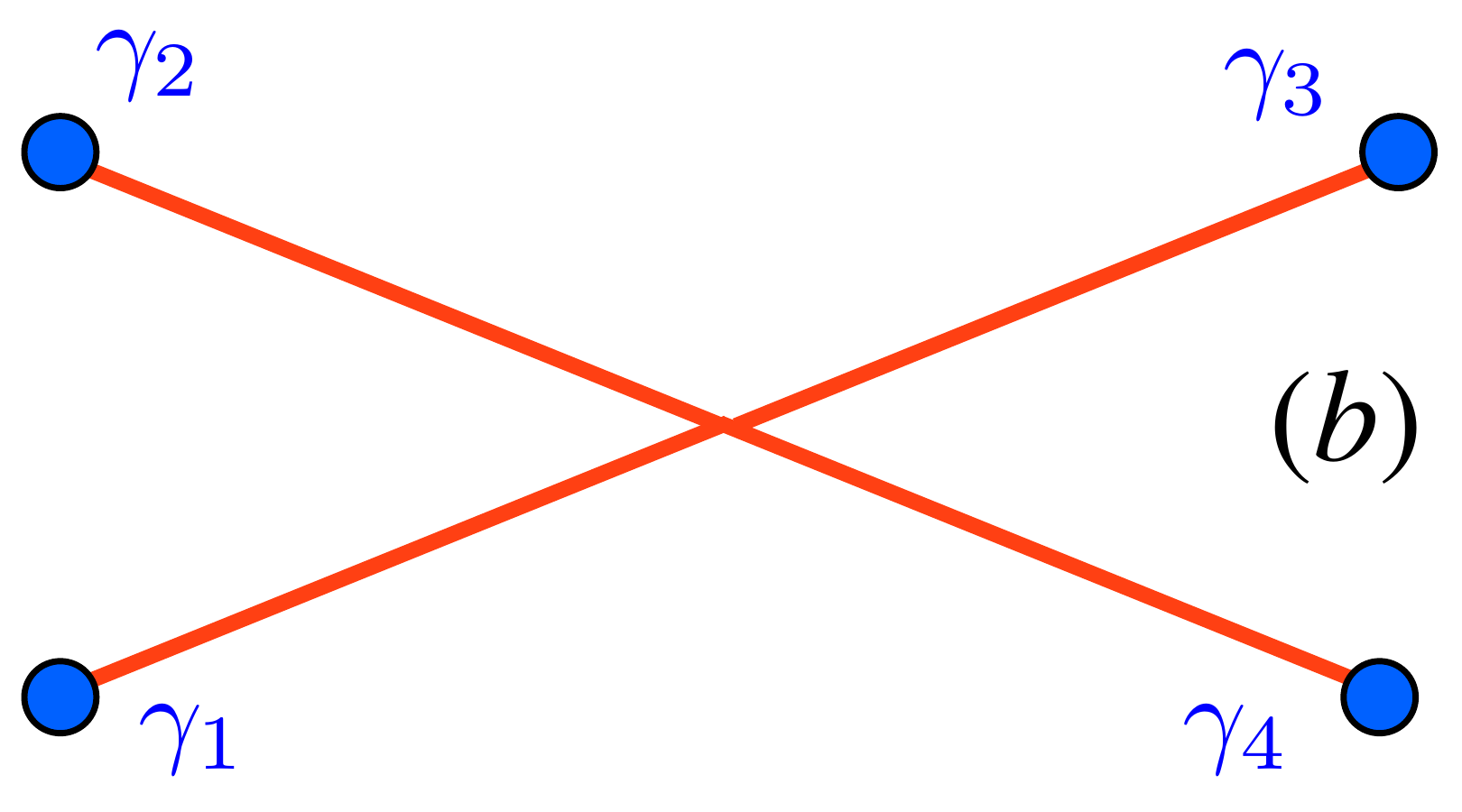} 
\caption{\label{fig:Xjunction}   Two possible realizations of the X-junction setup described by the Hamiltonian \eqref{eq:ham}. 
The panel (a) assumes four TI wires with endpoint Majorana states, four of them, $\gamma_{i}'$, are  far apart and four in the center, $\gamma_{i}$, are close enough to be described by the  hopping Hamiltonian \eqref{eq:ham}. If no special condition on this Hamiltonian is imposed, then four central Majorana states are fully split and disappear from the discussion, as shown in panel (b). The Hamiltonian \eqref{eq:ham} now refers to the remaining four Majorana states. 
  }
\end{figure}

We consider a system of four  connected wires with Majorana end modes, shown in Fig.\ref{fig:Xjunction}a,
that host eight  Majorana states.
However
the interaction between  four Majorana fermions at the center 
is parametrically stronger than with the Majorana states at the ends and should be considered at the first stage, as described for  Y-junction in Appendix C of Ref. \cite{Karzig2015}. 
If there are no special symmetries imposed on the system the degeneracy of the central states is  completely lifted,  and the remaining degrees of freedom are four weakly coupled Majorana states $\gamma_j'$.
Their  Hamiltonian (after suppressing prime symbol) is given by 
\begin{equation}
\label{eq5}
 {\cal H} = \tfrac i2\sum_{kl} H_{kl} \gamma_{k} \gamma_{l}
\end{equation}
 where 
 \begin{align}
H &  = \begin{pmatrix} 
 0  ,& u_{1} + \bar u_{1} , & u_{2} + \bar u_{2} , & u_{3} + \bar u_{3} \\ 
 \cdot & 0 ,&  -u _{3} + \bar u_{3} , & u_{2}  - \bar u_{2} \\ 
 \cdot   & \cdot  & 0, & -  u_{1} + \bar u_{1} \\ 
  \cdot   & \cdot  &  \cdot   & 0 
\end{pmatrix}
\label{eq:ham}
\end{align}
is  a  skew-symmetric matrix. 
The  special  case  of pairwise equality ($u_{j} = \bar u_{j}$) corresponds  to Eq.\ \eqref{eq:Yjunction}. 

Of course one cannot rule out the accidental symmetry which may result  in a setup
where  six Majorana states remain gapless. Though this situation is possible from a mathematical  point of view, it is highly unlikely to occur, unless it is specially tuned. 
Therefore we assume that the nearby Majorana states at the center of the junction are  gapped,  
while four  at the far ends remain gapless.
Such tuning can be accomplished by applying external gates and magnetic flux, as  
was discussed in details in Refs. \cite{Halperin2012,Romito2012,Hyart2013}. 
Clearly,  the distance between Majorana states  restricts a minimal  size of the gates.
It is therefore  reasonable   to  assume that the  larger distance between end Majoranas,  $\gamma'_{j}$ in Fig.\ \ref{fig:Xjunction}a, allows some  flexibility for changing the parameters in Eq.\  \eqref{eq:ham}.

Because  matrix $H$ is skew symmetric, it can be decomposed into  a sum
\begin{equation}
H=\sum_{j=1}^{3} (u_{j}g_{j} + \bar u_{j} \bar g_{j}),
\end{equation} 
where $g_{1,2,3}$ ($\bar g_{1,2,3}$ ) are the  generators of right (left) isoclinic subgroup of  $SO(4)$ group,
with the  standard  commutation relations $[g_{i},g_{j}] = 2\epsilon_{ijk} g_{k}$ , $[\bar g_{i}, \bar g_{j}] = -2\epsilon_{ijk} \bar g_{k}$. 
 Importantly, the generators of left and right isoclinic subgroup  commute,
$ [g_i,\bar{g}_j]=0\,. $  
The convenience of the above decomposition of $H$ becomes evident after performing  a unitary transformation, 
\begin{equation}
H \to U H U ^{-1}, 
\end{equation}
with $U$ belonging to $SO(4)$ group. It is known that any $U$ can be represented as a product $U= U_{L} U_{R} $ with 
\begin{eqnarray}&&
U_{R}= \exp\Big(\sum  _{j=1}^{3} \alpha_{j}g_{j} \Big) , \quad
U_{L}= \exp\Big(\sum _{j=1}^{3} \bar \alpha_{j} \bar g_{j} \Big).
\end{eqnarray}
 Since $g_{j}$ and $\bar g_{j}$ commute,  one finds
\begin{equation} 
 U H U ^{-1} = \sum_{j=1}^{3} (u_{j}  U_{R} g_{j} U_{R}^{-1}+ \bar u_{j} U_{L} \bar g_{j} U_{L} ^{-1}) .
\end{equation}
This decomposition  implies that  the triples $(u_{1}, u_{2}, u_{3})$ and $(\bar u_{1}, \bar u_{2}, \bar u_{3})$ are transformed within themselves according to   $SO(3)$ group,  that preserves   the ``lengths'' of the vectors $u = \sqrt{u_{1}^{2}+u_{2}^{2}+u_{3}^{2}}$ and $\bar u = \sqrt{\bar u_{1}^{2}+ \bar u_{2}^{2}+ \bar u_{3}^{2}}$ invariant. 
One can easily show that 
the  eigenvalues of $ {\cal H} $ are  $\pm (u+\bar u)/2$ and  $\pm (u-\bar u)/2$.
Because eigenvalues are preserved by  unitary transformation,  so are  
the  values of   $u$ and $\bar u$.
Therefore, for the case where  $u=\bar u$   two  Majorana states  (out of four) 
remain degenerate  even though  the components  $u_{j}$, $\bar u_{j}$ are not equal. 
 
Assuming that there are  two non-split Majorana states without loss of generality we assume  
$u=\bar u= 1$. The parametrization of   the vectors $u$ and  $\bar u$ by the angles \eqref{eq:param}, 
represents the Hamiltonian  in  new coordinates   $(\theta , \phi,  \bar \theta, \bar \phi )$ that 
correspond to two points on a unit sphere.
For a pairwise equality $u_{k}= \bar u_{k}$, $k=1,2,3$  the two points merge to one, reproducing  the familiar form of Y-junction Hamiltonian\eqref{eq:Yjunction}.

Now we employ this  parametrization to  calculate  the Berry's phase acquired by wave-function during the adiabatic evolution. Note, that the concept of adiabatic evolution in the present case needs a special justification,  due to  two-fold degeneracy of the zeroth-energy state. 
One may formally  justifies the procedure  by  infinitesimal  deformation  $u=1+\epsilon$ ,  $\bar u=1-\epsilon$, and finally sending   $\epsilon \to +0$. 
The adiabatic evolution of the  state  with the energy $\epsilon$,  is determined  
by the linear combination $c^{\dagger} =\sum_{k=1}^{4} A_{k} \gamma_{k}$,  where   
\begin{equation}
\begin{aligned}
\mathbf {A}=   &  \tfrac1{2}\big( 
-\sin \theta _+ \sin \phi_-   -i \sin \theta _-  \cos \phi _- ,  \\ 
 &  -\sin \theta _+    \cos  \phi _- +i \sin  \theta_-  \sin \phi _- ,    \\ 
  &  \cos   \theta _+  \cos \phi _+ +i   \cos \theta _-  \sin  \phi   _+ ,   \\
  & \cos  \theta _+  \sin \phi _+ -i \cos \theta _-    \cos  \phi _+ \big)  .
\end{aligned}
\label{def:A}
\end{equation}

Here we used the notations $|\mathbf {A}|^{2} = \langle A| A \rangle=1/2$, and 
$ \theta _\pm  = (\theta \pm  \bar \theta )/2   , \quad 
    \phi _\pm   = ( \phi \pm  \bar \phi )/2$. 

Clearly, the  state with the energy $-\epsilon$ is given by $c =\sum_{k=1}^{4} A^{*}_{k} \gamma_{k}$. 
The geometrical phase, $\alpha$,  acquired by the wave function during  the evolution is 
given by
\begin{equation}
\frac{d\alpha}{dt} = \langle A | i\frac{d}{dt}  | A \rangle 
= -\frac 14  \left(\cos \theta \frac{d\phi}{dt} + \cos \bar\theta \frac{d\bar \phi}{dt} \right).
\end{equation}
Let us now assume that  evolution follows the closed trajectory and the final configuration is the same as the initial one, $(\theta , \phi,  \bar \theta, \bar \phi )$.
In this case,  the Berry's phase can be written as the integral over the closed path
\begin{equation}
\begin{aligned}
 2\alpha  =   
 -\frac 12  \oint \left(\cos \theta\, {d\phi} + \cos \bar\theta \,{d \bar \phi} \right) . 
\end{aligned}
\end{equation}
If  initially $\mathbf {A}$ is a linear combination of $\gamma_{3,4}$,  then 
the trajectory  starts from the north pole ($\theta =  \bar \theta =0$).  Using 
 Stokes'  theorem one  transforms the line integral  to the  area  integral 
\begin{equation}
\begin{aligned}
2\alpha= \frac 12  \int\sin \theta \,d\theta d\phi  + 
\frac 12 \int \sin \bar\theta \,d\bar\theta  d\bar \phi\,.
\end{aligned} 
 \label{eq:Berry2}  
\end{equation}
As we  see, the  geometric phase  is a sum of two contribution. 
In the limit of  $\theta = \bar \theta$, $ \phi = \bar \phi $  one reproduces 
Eq.\ \eqref{eq:Berry1}.  
In general case, however,  there is a large freedom in choosing the trajectories, that  
are consistent with quantum computations. This allows us to generate almost Berry's phase with any values.
We will now demonstrate two simple examples.

At first we discuss a setup where the condition   $\theta = \bar \theta$  is maintained during the entire cycle.
Examining  Eq.\eqref{eq:Berry2},  we see that 
the corresponding Berry's phase  
is essentially given by the previous formula \eqref{eq:Berry1} with the replacement $\phi\to \phi_{+} = ( \phi +  \bar \phi )/2$. 
It means that in this type of evolution  the difference $\phi_{-}$ between $\phi$ and $\bar{\phi}$ plays no role,
and does not  contribute to Berrys' phase.  This property might prove to be useful in error correction protocols such as, e.g., proposed  in \cite{Karzig2015}.


In the second protocol we consider,  the value of   $\bar \phi$ is fixed,  and  the overall Berry's phase   is given only by the first integral in \eqref{eq:Berry2}, which is one half of the expression \eqref{eq:Berry1}. 
In particular,  the desired value of $\alpha=\pi/8$ is 
provided by the previous trajectory around the octant of the sphere, Fig. \ref{fig:sphere}, 
explicitly defined by  $(\theta, \phi): (0,0) \to (\pi/2, 0)\to (\pi/2, \pi/2)\to (0,\pi/2)$. 
The endpoints of the trajectory in second pair of coordinates $(\bar \theta, \bar \phi)$  should corresponds to the north pole, $\bar\theta=0$. The intermediate values of $\bar \theta$ are unimportant, and this   
 additional degree of freedom  may be helpful in discussing possible ways to reduce the errors in $\alpha$, see  Ref.\cite{Karzig2015}. 
 
The explicit form of the X-junction Hamiltonians, corresponding to two  cases above,
is shown  in Appendix \ref{sec:App}.



\section{Conclusions and outlook}
In this paper we studied the manipulation of  Majorana fermions coupled by X-junction. 
Unlike a more conventional Y-junction geometry, this setup requires a special tuning, 
in order to maintain the  degeneracy of its  ground state. 
The bright side of this method is  a big  parameter space that allows to perform  necessary  unitary transformations.
Within this space  there is  a  surface, where two Majorana bound states remain non-split. 
This surface is  preserved  under  a group of unitary transformations, 
that factorizes into right and left isoclinic groups.
Provided that  the coupling of the junctions are changed in accordance with this symmetry 
the acquired phase is unaffected by the dephasing, and the junction can be used as a quantum gate.
We proposed  a few examples of adiabatic manipulations, that are consistent with this requirement,  and calculate the corresponding Berry's phase. In particular,  the trajectories generating the vlaue $\alpha= \pi/8$, as well as other simple  fractions of $\pi$, are demonstrated.

The  natural  extension of  this analysis are  junctions with five and more connectors.
In this case, the system has a higher symmetry,  and one expects more  ways to perform the  dephasing-free unitary transformations.


Aiming at  applications,
we understand that in possible realization the dephasing due to noise in the control gates could be quite substantial so that the X-junction would not be very different from any other non-topological qubit. The previously discussed Y-junction configuration is not immune to decoherence, which enters mainly via the $\pi/8$ gate. Thus it would be interesting to know if the X-junction, with generally absent topological protection,  could compete in this respect with the Y-junction geometry.
Particularly it is yet to be seen, whether one can design a physical realization,
that locks the junction on to dephasing-free surfaces on the hardware level. 
Since the couplings in the Hamiltonian are influenced by external noise, 
it implies  a design where noise affects a few couplings in a correlated manner. 
Unfortunately at this point, we were not able to propose a design based on 
realistic elements, where this locking takes place. 
However, even though a topologically protected quantum computation remains evasive, 
it may happen that dephasing time for a particular device 
will be sufficiently long for application. Our work lays a mathematical foundation
for constructing quantum gates via X-junction.

\acknowledgments
We thank Yuval Oreg for useful discussions.
This work was supported by the RFBR grant No 15-52-06009 and by GIF (grant 1167-165.14/2011).
D.G. acknowledges the  support by ISF (grant 584/14) and 
Israeli Ministry of Science, Technology and Space.  
 
 \appendix* \section{}
 \label{sec:App}
In this appendix we present an explicit form for the effective Hamiltonian that correspond to the protocols
suggested above. For the case $\theta=\bar \theta$ the Hamiltonian is  given (up to the overall scale) by
\begin{eqnarray} 
H  &  =& \begin{pmatrix} 
 0  ,& a , & b_{1}, &  b_{2}\\ 
 \cdot & 0 ,& b_{3}, &  b_{4}\\ 
 \cdot   & \cdot  & 0, & 0 \\ 
  \cdot   & \cdot  &  \cdot   & 0 
\end{pmatrix} , 
\label{eq:ham3} 
\end{eqnarray}  
 where we define
\begin{equation*}
 \begin{aligned}
  a &= \cos \theta  ,  \\
b_{1} & = \sin \theta  \cos  \phi_{-} \cos\phi_{+} ,  \quad
b_{2}  = \sin \theta \cos \phi_{-} \sin\phi_{+} ,     \\ 
b_{3} & = -\sin \theta \sin\phi_{-} \cos \phi_{+}  ,  \quad 
b_{4}  =- \sin \theta \sin\phi_{-}  \sin\phi_{+}.     
 \end{aligned}
\end{equation*}  
This shows that the hopping amplitude between $\gamma_{3}$ and $\gamma_{4}$ is zero during the whole cycle. 

In the second protocol discussed after Eq. \eqref{eq:Berry2} the value of   $\bar \phi$ is fixed, and we set $\bar \phi=\pi/2$ for definiteness.  We denote $c =\cos \bar \theta$, $s=\sin \bar \theta$ for brevity, then the   Hamiltonian acquires the form 
\begin{equation}
H =   \begin{pmatrix}
 0 &c + \cos \theta  &  \sin \theta  \cos \phi   & s + \sin \theta  \sin \phi  \\
\cdot & 0 & s-\sin \theta  \sin   \phi  &  \sin \theta \cos \phi  \\
\cdot & \cdot & 0 & c-\cos \theta  \\
 \cdot & \cdot  & \cdot & 0 \\
\end{pmatrix}.
\label{eq:ham2setup}
\end{equation}
Notice that \eqref{eq:ham2setup} allows for non-vanishing coupling between $\gamma_{3}$ and $\gamma_{4}$ during  manipulation.   The endpoints of the trajectory should correspond to  $c=1$, $s=0$, which is the north pole in coordinates $(\bar \theta, \bar \phi)$. The intermediate values of $c$, $s$ are unimportant for Berry's phase.


\end{document}